\documentclass[twocolumn,aps,showpacs,floatfix,superscriptaddress]{revtex4}

\usepackage{graphicx}
\usepackage{revsymb}

\begin{document}
\title{Evaluation of the total photoabsorption cross sections for actinides from photofission data
and model calculations}
\author{I.A. Pshenichnov}\email[e-mail: ]{pshenichnov@AL20.inr.troitsk.ru}
        \affiliation{Center for Nuclear Studies, Department of Physics,
                     The George Washington University, Washington, D.C. 20052, USA}
        \affiliation{Institute for Nuclear Research, Russian Academy of Science,
                      117312 Moscow, Russia}
\author{B.L.~Berman}\email[e-mail: ]{berman@gwu.edu}
        \affiliation{Center for Nuclear Studies, Department of Physics,
                     The George Washington University, Washington, D.C. 20052, USA}
\author{W.J.~Briscoe}\email[e-mail: ]{briscoe@gwu.edu}
        \affiliation{Center for Nuclear Studies, Department of Physics,
                     The George Washington University, Washington, D.C. 20052, USA}
\author{C.~Cetina}\altaffiliation[Present address: ]{Naval Research Laboratory, Washington, D.C. 20375, USA}
        \affiliation{Center for Nuclear Studies, Department of Physics,
                     The George Washington University, Washington, D.C. 20052, USA}
\author{G.~Feldman}\email[e-mail: ]{feldman@gwu.edu}
        \affiliation{Center for Nuclear Studies, Department of Physics,
                     The George Washington University, Washington, D.C. 20052, USA}
\author{P.~Heimberg}\altaffiliation[Present address: ]{Bechtel-Nevada, Andrews Air Force Base, MD 20762, USA}
        \affiliation{Center for Nuclear Studies, Department of Physics,
                     The George Washington University, Washington, D.C. 20052, USA}
\author{A.S.~Iljinov}
         \affiliation{Institute for Nuclear Research, Russian Academy of Science,
                      117312 Moscow, Russia}
\author{I.I.~Strakovsky}\email[e-mail: ]{igor@gwu.edu}
        \affiliation{Center for Nuclear Studies, Department of Physics,
                     The George Washington University, Washington, D.C. 20052, USA}
\begin{abstract}
We have calculated the fission probabilities for $^{237}{\rm Np}$,
$^{233,235,238}{\rm U}$, $^{232}{\rm Th}$, and $^{\rm nat}{\rm
Pb}$ following the absorption of photons with energies from 68 MeV
to 3.77 GeV using the RELDIS Monte-Carlo code. This code
implements the cascade-evaporation-fission model of
intermediate-energy photonuclear reactions. It includes
multiparticle production in photoreactions on intranuclear
nucleons, pre-equilibrium emission, and the statistical decay of
excited residual nuclei via competition of evaporation, fission,
and multifragmentation processes. The calculations show that in
the GeV energy region the fission process is not solely
responsible for the entire total photoabsorption cross section,
even for the actinides. The fission probabilities are  $80-95$\%
for $^{233}{\rm U}$,$^{235}{\rm U}$, and $^{237}{\rm Np}$,
$70-80$\% for $^{238}{\rm U}$, and only $55-70$\% for $^{232}{\rm
Th}$. This is because certain residual nuclei that are created by
deep photospallation at GeV photon energies have relatively low
fission probabilities. Using the recent experimental data on
photofission cross sections for  $^{237}{\rm Np }$ and
$^{233,235,238}{\rm U}$ from the Saskatchewan and Jefferson
Laboratories and our calculated fission probabilities, we infer
the total photoabsorption cross sections for these four nuclei.
The resulting cross sections per nucleon agree in shape and in
magnitude with each other. However, disagreement in magnitude with
total-photoabsorption cross-section data from previous
measurements for nuclei from C to Pb calls into question the
concept of a ``Universal Curve'' for the photoabsorption cross
section per nucleon for all nuclei.
\end{abstract}
\maketitle

\section{Introduction}

\subsection{Relevant Framework}

Recently, high-precision experimental data have been obtained for
the photofission of actinide and preactinide nuclei at the
Saskatchewan Accelerator Laboratory (SAL) and Jefferson Laboratory
(JLab)~\cite{Sanabria,SanabriaPhd,Cetina,Cetina2002}. These data
constitute both a challenge and an opportunity: a challenge to
theory to reproduce them, and an opportunity to use them to
determine the total photoabsorption cross sections for these
nuclei. From these one can throw light, for example, on the
concept of a ``Universal Curve'' for the photoabsorption cross
section per nucleon for all nuclei~\cite{Ahrens1984}.

At this time, the only theory capable of reproducing these data is
an extension of the widely-known Intranuclear Cascade (INC)
model~\cite{CRC} to GeV photon-induced reactions~\cite{Iljinov},
including subsequent evaporation and fission. Almost three decades
ago, the INC model was successful in describing experimental data
obtained with bremsstrahlung photons~\cite{Barashenkov}. Later,
the hybrid precompound-evaporation model approach of
Ref.~\cite{Blann1983} was successful in describing multiple
photoneutron reactions below the pion-production threshold.

A related subject is the electromagnetic dissociation of
relativistic heavy ions, where virtual photons are used to
initiate photonuclear reactions~\cite{Heckman,Olson}. Recently,
the INC model~\cite{Iljinov} has been applied to the
electromagnetic dissociation of ultrarelativistic heavy
ions~\cite{Pshenichnov,Pshenichnov2,Pshenichnov2001,Scheidenberger}.
In such reactions, nuclei are disintegrated by virtual photons
over the wide energy range from a few MeV to a few tens of GeV. A
Monte-Carlo code called RELDIS (Relativistic ELectromagnetic
DISsociation) was devised to perform calculations for real and
virtual photons across this wide energy range.

We now use the RELDIS code to describe the recent photofission
data~\cite{Sanabria,SanabriaPhd,Cetina,Cetina2002}. This paper
reports our results and their implications. The description of the
RELDIS model of photonuclear reactions is given in
Sec.~\ref{RELDIS_descr}. Calculational results for nuclear fission
probabilities and absolute photofission cross sections are
presented in Sec.~\ref{Res_Disc}. Evaluation of the
photoabsorption cross section per bound nucleon, commonly referred
to as the ``Universal Curve,'' from the actinide photofission data
and calculated fission probabilities is presented in
Sec.~\ref{Res_Disc}. We summarize our findings in
Sec.~\ref{Concl}.

\subsection{Experimental Background}

As first pointed out by Bohr and Wheeler~\cite{Bohr-Wheeler} and
echoed by Aage Bohr~\cite{Bohr}, the relative simplicity and
directness of the electromagnetic interaction is very useful in
the exploration of the process of nuclear fission. At the first
stage of a photonuclear reaction, the absorption of a photon on a
pair of intranuclear nucleons brings in only thermal energy and
produces fewer changes in the structure of the target nucleus
than, for example, reactions induced by protons or antiprotons. In
the latter cases, the antiproton annihilation removes an
intranuclear nucleon, or, respectively, the projectile nucleon can
be trapped by the nuclear potential. Therefore, one can expect
that following photoabsorption at low energies, most of the
fissioning nuclides are similar to the target nucleus. This makes
photofission studies much more transparent.

Over the years, photon-induced fission has attracted attention,
and considerable progress has been made in experimental
photofission studies. In the Giant Dipole Resonance (GDR) region,
experiments at Livermore~\cite{Caldwell,Caldwell1,Caldwell2} and
elsewhere delineated the giant-resonance parameters with high
accuracy for eight actinide isotopes ($^{232}{\rm  Th}$,
$^{237}{\rm Np}$, $^{239}{\rm Pu}$, and the five long-lived
uranium isotopes). At these low energies (from threshold to about
20 MeV), photofission reveals its simplicity, and the total
photoabsorption cross section equals the sum of the single- and
double-photoneutron cross sections plus the photofission cross
section, since the high Coulomb barrier greatly inhibits the
emission of charged particles.

In the quasideuteron and $\Delta_{33}(1232)$ regions, experiments
at Saclay~\cite{Lep-fiss}, Mainz~\cite{Frommhold1992}, and
elsewhere extended the photofission data to intermediate energies
for $^{235,238}{\rm U}$ and $^{232}{\rm Th}$. From a comparison of
the photofission cross sections per nucleon for these nuclei with
the total photoabsorption cross sections per nucleon obtained by
other methods at Mainz~\cite{Frommhold1994},
Frascati~\cite{Bianchi}, and Bonn~\cite{Valeria}, it was commonly
believed that there was a ``Universal Curve'' and that the
photofission cross section for these nuclei saturates the total
cross section. The recent precise data from
SAL~\cite{Sanabria,SanabriaPhd} agree well with the earlier data
from Saclay and Mainz across the upper part of the quasideuteron
region (about 60 MeV to the photopion threshold) and the lower
part of the $\Delta_{33}(1232)$ region (up to about 250 MeV).

The recent data from JLab~\cite{Cetina,Cetina2002} extend our
knowledge of the photofission cross sections for $^{237}{\rm Np}$,
$^{233,235,238}{\rm U}$, $^{232}{\rm Th}$, and $^{\rm nat}{\rm
Pb}$ across the $\Delta_{33}(1232)$ region and through the
higher-resonance region (from 0.20 to 3.77 GeV). In this energy
domain, photonuclear reactions seem to be more complicated. New
advanced theoretical models must be employed to account for the
greater number of reaction channels that are open during the first
step of the reaction due to meson photoproduction on the
intranuclear nucleons. Also, the understanding of the role of
nuclear fission among the many other decay modes of the excited
nucleus requires a well-founded theoretical model to describe such
decay.

\subsection{Theoretical Approaches to Describe Photofission}\label{SC1}

One of the first calculations using the Intranuclear Cascade (INC)
model that includes photon absorption at intermediate energies was
that of Ref.~\cite{Barashenkov}. This calculation takes into
account the channels of the $\gamma N$ interaction with production
of only one or two pions, and thus can be applied at photon
energies only up to 1~GeV. It was tested with old experimental
data obtained with bremsstrahlung photons. Later, the model was
successfully applied to early photofission
studies~\cite{Cherepanov,Iljinov-Mebel,Guaraldo}.

INC model predictions~\cite{Barashenkov,Iljinov-Mebel} for the
first pre-equilibrium stage of the photofission reaction were also
used in fission probability calculations for $^{232}{\rm Th}$ and
$^{238}{\rm U}$ nuclei by introducing the mean-compound-nucleus
approximation~\cite{Arruda-Neto}. According to this method, the
ensemble of excited residual nuclei created after the first step
of a photonuclear reaction is replaced by a single excited nucleus
with the average values of neutrons $N$, protons $Z$, and
excitation energy $E^\star$. In order to explain the data on
thorium, one needs to account for its higher nuclear transparency
relative to uranium~\cite{Arruda-Neto}. Such a difference for
nuclear systems of comparable mass was attributed to subtle
details of the nuclear structure of $^{232}{\rm Th}$ and
$^{238}{\rm U}$. The nature and origin of such differences still
need to be explained. The fissility of $^{238}{\rm U}$ was assumed
to be saturated; that is, its fission probability $W_f\equiv
\sigma_f/\sigma_{tot}$ was assumed to be unity.

Another phenomenological method was proposed in
Ref.~\cite{Delsanto} to describe the photofission of $^{209}{\rm
Bi}$, $^{232}{\rm Th}$, and $^{238}{\rm U}$ in the quasideuteron
region, $E_\gamma=30-140$~MeV. A special phenomenological factor,
which selects the quasideuteron absorption that leads to
photofission, was introduced and was found to be different for
$^{209}{\rm Bi}$ than for $^{238}{\rm U}$.

In Refs.~\cite{Deppman:2001,Deppman:2002} the photofissility of
actinide nuclei at intermediate energies was considered within the
multicollisional model for photon-induced intranuclear cascade
process and the statistical neutron evaporation and fission models
for deexcitation of residual nuclei. The fission probabilities for
both $^{237}{\rm Np}$ and $^{238}{\rm U}$ were found to be
unsaturated, {\it i.e.} very close, but still less than unity.
However, the authors of Refs.~\cite{Deppman:2001,Deppman:2002}
considered their approach to be accurate only in a restricted
range of photon energies, from 0.5 GeV to 1 GeV. This makes
difficult direct comparison with the SAL and JLab
data~\cite{Sanabria,SanabriaPhd,Cetina,Cetina2002}, since the data
were obtained over a much wider energy range, from 68 MeV to 3.77
GeV.

Recently, the range of applicability of  the INC model was
extended up to 10~GeV~\cite{Iljinov}. The accuracy of the model
was improved from the photoneutron thresholds up to 2~GeV, and new
data obtained with monochromatic photons were used to verify the
model predictions. The analysis of photofission reactions
performed in the present paper is based on this approach, which is
free of the simplifications and assumptions made in
Refs.~\cite{Arruda-Neto,Delsanto} and is valid for a much wider
region of incoming photon energies than the approach of
Refs.~\cite{Deppman:2001,Deppman:2002}.

The following questions are addressed here:
\begin{itemize}

\item[(1)] Is it possible to describe the photofission of actinide nuclei over such a broad energy range, and
           if so, how good is this description?
\item[(2)] Does the relative simplicity of low-energy photonuclear reactions persist
           at photon energies above 1~GeV, where multiple photoproduction
           of hadrons takes place?
\item[(3)] To what extent does the excited residual nucleus, created after
           the first stage of the photonuclear reaction, retain the properties
           of the target nucleus?
\item[(4)] Can the total photoabsorption cross section for heavy actinide nuclei
           be obtained from the total photofission cross section?
\item[(5)] Is the concept of the ``Universal Curve,'' which has been obtained for $A\leq 208$ nuclei,
           also valid for $A\geq 233$ nuclei?
\item[(6)] What is the level of calculational and experimental uncertainties in the evaluation of
           the total photoabsorption cross section for heavy actinides?
\end{itemize}
\noindent In order to answer these questions, we use the model
with its main parameters extracted from independent
studies~\cite{Iljinov,Pshenichnov,Pshenichnov2,Cherepanov,Iljinov-Mebel,golu94,bot90,JPB}
of photon-, hadron-, and heavy-ion-induced reactions, in contrast
to other
approaches~\cite{Arruda-Neto,Delsanto,Deppman:2001,Deppman:2002},
where only photon-induced reactions were considered. The present
theoretical investigation is aimed at estimating the
characteristics of the first step in the photoabsorption process
and determining the total fission probability for such a reaction.

\section{The RELDIS model of Photonuclear Reactions}\label{RELDIS_descr}

\subsection{Initial Interaction and Intranuclear Cascade of Photoproduced
Hadrons}

The fast hadrons produced in a primary $\gamma N$ or $NN$
interaction initiate a cascade of successive hadron-nucleon
collisions inside the target nucleus during the first,
nonequilibrium, stage of the photonuclear reaction. The duration
of this stage $\tau_{cas}$ can be estimated as the time a fast
particle needs to cross the nucleus: $\tau_{cas}\sim\tau_o$, where
$\tau_o\leq 10^{-22}$ s.

The INC model is a numerical method to solve the kinetic equation
that describes hadron transport in the nuclear medium.
Calculations of intranuclear cascades were performed by using a
Monte Carlo technique. The nucleus is considered to be a mixture
of degenerate Fermi gases of neutrons and protons in a spherical
potential well with a diffuse boundary. By using the effective
real potentials of nucleons and pions, the influence of
intranuclear nucleons on these cascade particles is taken into
account. The momentum distribution of intranuclear nucleons is
calculated in the local-density approximation of the Fermi gas
model. The distribution of nuclear density is approximated by a set
of step-like functions. The cross sections of elementary collisions
$NN\rightarrow NN$, $NN\rightarrow \pi NN$, $\pi N\rightarrow \pi
N$, $\pi N\rightarrow \pi\pi N$,  $\pi N N\rightarrow N N$,$\ldots$
in the nuclear medium are assumed to be the same as in vacuum
except that using the Pauli Principle prohibits transition of the
cascade nucleons into the states already occupied by the
intranuclear nucleons. A more detailed description of the INC
model is given in Ref.~\cite{CRC}.

By means of the INC model, the process of dissipation of the
initial photon energy can be investigated in detail. Part of this
energy is transformed into internal excitation of the residual
nucleus; the rest is released by fast cascade hadrons leaving the
nucleus.

\subsection{Creation of an Excited Compound Nucleus}\label{INC}

At the end of the hadronic cascade, some quasiparticles remain in
the nuclear Fermi gas as ``holes'' ($N_h$), which are knocked-out
nucleons, and particles  ($N_p$), which are slow cascade nucleons
trapped by the nuclear potential. In the INC model, the excitation
energy of the residual nucleus is defined as
\begin{equation}
E^\star=\sum_{i=1}^{N_p}\epsilon^p_i+\sum_{i=1}^{N_h}\epsilon^h_i,
\end{equation}
where the quasiparticle energies for particles $\epsilon^{p}_i$
and holes $\epsilon^{h}_i$ are measured with respect to the Fermi
energy. The numbers of nucleons and protons of the residual
nucleus are given by the relations
\begin{equation}
A_{RN}=A-\sum_{i=1}^{N_c}q^c_i
\end{equation}
\noindent and
\begin{equation}
Z_{RN}=Z-\sum_{i=1}^{N_c}e^c_i,
\end{equation}
where $A$ and $Z$ are the numbers of nucleons and protons,
respectively, in the target nucleus, and $q^c_i$ and $e^c_i$ are
the baryon number and charge carried away by the cascade particle
$i$.

The moment when all fast particles have left the nucleus marks the
beginning of the period of the establishment of thermal
equilibrium in the residual nucleus with a duration of
$\tau_{peq}\sim (10-100)\cdot \tau_o$. At the end of this period,
thermal equilibrium is reached and an excited compound nucleus has
been created.

Various criteria may be applied to decide whether the nuclear
system has reached thermal equilibrium. According to the exciton
model~\cite{Toneev}, the system of Fermi particles created after
the cascade stage is not in equilibrium if the number of
quasiparticles $N_q=N_p+N_h$ is less than the equilibrium value
$N^{eq}_q\sim \sqrt{2gE^\star}$, where $g$ is the density of
single-particle states. Therefore, in the course of the
equilibration process, pre-equilibrium particles may be emitted.
In our calculations, the pre-equilibrium exciton
model~\cite{Toneev} is used to simulate pre-equilibrium emission.

Relative probabilities of the de-excitation processes are
determined by the excitation energy $E^\star$ and by the mass
$A_{RN}$ and charge $Z_{RN}$ of the residual nucleus formed after
the establishment of thermal equilibrium.

Depending on the photon energy $E_\gamma$, various processes may
contribute to the energy deposition. In the following, we consider
step-by-step all of these mechanisms, with special attention to
the amount of energy that is transformed into internal excitation
of the nucleus.

\begin{figure}[ht]
\begin{centering}
{\includegraphics[width=0.85\columnwidth]{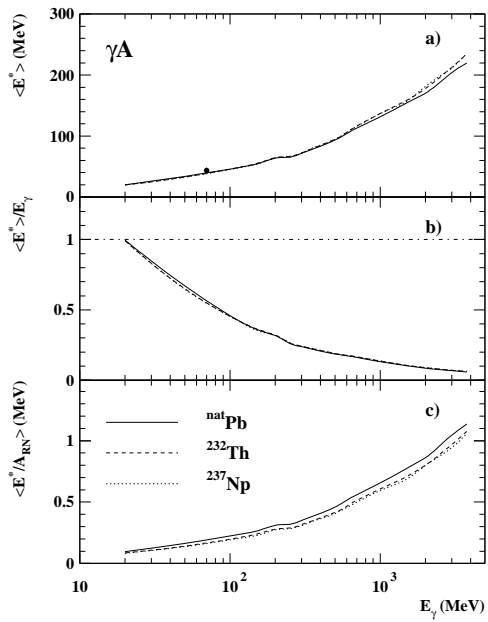}}
\caption{a) Average excitation energy of the residual nucleus
$\langle E^\star\rangle $ following photoabsorption.
b) Ratio of the average excitation energy of the residual nucleus
$\langle E^\star\rangle $
to the input photon energy $E_\gamma$.
c) Average values of $E^\star$ per nucleon of the residual nucleus.
The RELDIS code results are given by solid, dashed, and dotted lines
for $^{\rm nat}{\rm Pb}$, $^{232}{\rm Th}$, and $^{237}{\rm Np}$,
respectively.
The value for $\langle E^\star\rangle$
deduced from an experiment on the photoabsorption for
lead~\protect\cite{Lepretre3} is denoted by the data point.
}
\label{fig:0}
\end{centering}
\end{figure}

Figure~\ref{fig:0} shows the average excitation energy $E^\star$
and the fraction of $E_\gamma$ that on average is transformed into
$E^\star$ in photoabsorption on $^{\rm nat}{\rm Pb}$, $^{232}{\rm
Th}$, and $^{237}{\rm Np}$. The average values of $E^\star$ per
nucleon of the residual nucleus $\langle E^\star/A_{RN}\rangle$
also are shown in Fig.~\ref{fig:0}.

%% Fig. 1
%%

As shown by calculations, when a nucleus absorbs a photon in the
GDR region ($6\leq E_\gamma \leq 30$~MeV) its energy is almost
completely transformed into excitation energy $E^\star$. For a
preactinide nucleus like $\rm Pb$, whose fission threshold is
above 20~MeV, the de-excitation proceeds mainly through the
evaporation of neutrons, since their separation energies are only
about 7~MeV. Due to the high Coulomb barrier in heavy nuclei,
proton emission is suppressed in the GDR region. Fission
thresholds are much lower for the actinides; {\it e.g.}, for U and
Np, it is $\leq 6$~MeV~\cite{Caldwell}, and the fission channel
dominates in the decay of such nuclei even at low excitation
energies~\cite{Caldwell,Caldwell1,Caldwell2}.

Starting from $E_\gamma =$ 30~MeV, where the quasideuteron
mechanism becomes important, up to the single-pion-production
thresholds at $E_\gamma\simeq 140$~MeV, only a small part of the
photon energy is converted (on average) into the excitation energy
$E^\star$ of the residual nucleus. The rest of the photon energy
is taken away by the fast nucleons originating from the absorbing
pair. It was deduced from experimental data~\cite{Lepretre3} that
$E^\star = 43.4\pm5$~MeV for photoabsorption on lead at
$E_\gamma=70$~MeV. The prediction of our model agrees well with
this value, as seen in Fig.~\ref{fig:0}a.

The two-nucleon absorption cross section of a photon on a heavy
nucleus $\sigma^{QD}_{\gamma A}$ is taken from the quasideuteron
model of Levinger ~\cite{Levinger}, as modified in
Ref.~\cite{Lepretre}:
\begin{equation}
\sigma^{QD}_{\gamma A} = k Z (1-Z/A) \sigma_d^{exch}.
\label{eq:13}
\end{equation}
\noindent Here $\sigma_d^{exch}$ is the meson-exchange part of the
cross section $\sigma_d$ for deuteron photodisintegration, $\gamma
d \rightarrow np$~\cite{Laget}, $A$ and $Z$ are the mass and
charge numbers of the relevant nucleus, and $k\approx$11 is an
empirical constant taken from the analysis of
Ref.~\cite{Lepretre}.

Although the cross section $\sigma_d$ decreases strongly with
photon energy, the two-nucleon absorption mechanism competes with
the single-nucleon photoabsorption channel $\gamma N\rightarrow
\pi N$ even up to $E_\gamma\sim 500$~MeV, when the wavelength of
the incident photon  becomes much smaller than the internucleon
spacing.

An interesting effect concerning the photon energy dissipation
above the $\gamma N\rightarrow \pi N$ threshold was noticed in
Ref.~\cite{Guaraldo}. A pion of 50--100 MeV has a small
interaction cross section with nucleons and therefore has a high
probability to carry away a large part ($\approx m_\pi$) of the
photon energy. Only at $E_\gamma\approx 250$~MeV does the average
value $\langle E^\star\rangle $ start to increase, as shown in
Fig.~\ref{fig:0}.

The reliability of the model predictions for $E^\star$ can be
tested by the comparison of the calculated first and second
moments of multiplicity distributions of neutrons, $\langle
N_n\rangle$ and $W_n=\sqrt{\langle N_n^2\rangle -\langle
N_n\rangle ^2}$, with the experimental data of
Ref.~\cite{Lepretre3}. In photoabsorption on lead, most of the
neutrons are emitted via evaporation from excited compound nuclei,
and both $\langle N_n\rangle$ and $W_n$ are sensitive to
$E^\star$. This test was performed in Ref.~\cite{Pshenichnov2} and
$\langle N_n\rangle$ and $W_n$  were found to be described with an
accuracy of 10-15\% at $70\leq E_\gamma\leq 140$ MeV.

Above the two-pion production threshold, at $E_\gamma\sim
400$~MeV, the photon-nucleon ($\gamma N$) interaction becomes more
complicated because of the opening of many possible final states.
We use a phenomenological model for the description of the $\gamma
N$ interaction, developed in Ref.~\cite{Iljinov}. The model
includes not only the excitation of nucleon resonances, but also
both the resonance contribution from the two-body channels,
$\gamma N \rightarrow \pi B^\star$ and $\gamma N \rightarrow
M^\star N$ ($B^\star$ and $M^\star$ being baryon and meson
resonances) and the nonresonant contribution from the
multiple-pion production channels $\gamma N \rightarrow i \pi N$
($2 \leq i \leq 8$). Finally, when the photon energy reaches a
value of a few GeV,  multiple-pion production becomes the dominant
process. A large number ($\sim 80$) of many-body subchannels are
included in our calculation.

Since the cross section $\gamma N\rightarrow hadrons$ is small
compared to the total $NN$ or $\pi N$ cross sections, this leads
to a large number of photons leaving the nucleus without producing
hadrons in the case of direct use of a Monte-Carlo simulation
technique. This is a natural consequence of the well-known fact
that nuclei are highly transparent to photons (i.e., photons do
not interact strongly). In order to reduce computation time, we
simulate photon absorption in each event, but in this case we have
to normalize our simulated results to the total photonuclear cross
section to get the absolute fission cross sections.

At high photon energies, many hadrons are produced inside the
target nucleus in the course of a cascade process. Due to the
knockout of intranuclear nucleons by fast nucleons and pions, the
slower particles pass through a lower-density region, thereby
undergoing fewer rescatterings; this is the so-called ``trawling''
effect. As was shown in Refs.~\cite{bot90,golu94,Lott}, the
trawling effect is important for a realistic description of
reactions at projectile energies above several GeV. In the present
calculation, we use a nonlinear version of the INC
model~\cite{golu94}, which takes into account the local depletion
of nuclear density during the development of the intranuclear
cascade.

\begin{figure}[ht]
\begin{centering}
{\includegraphics[width=1.1\columnwidth]{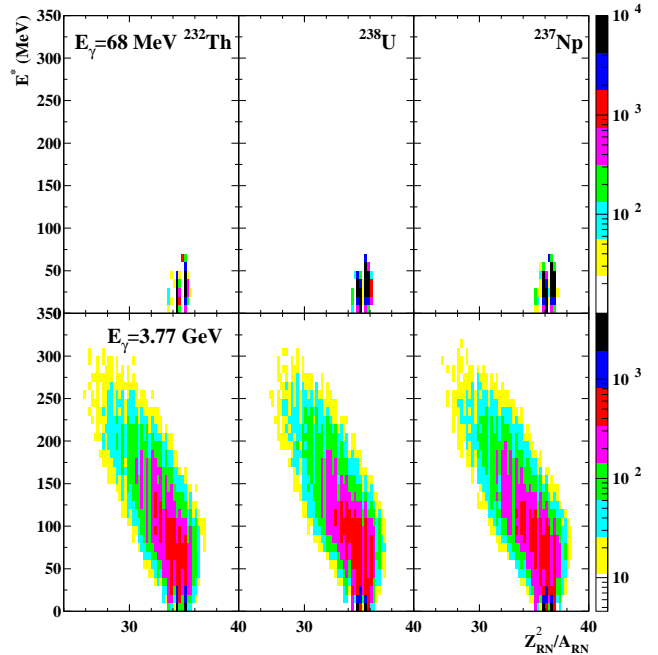}}
\caption{Distributions of $E^\star$ and $Z^2_{RN}/A_{RN}$ for
residual nuclei created after the fast stage of the
photoabsorption process for $E_\gamma=68$~MeV (top) and for
$E_\gamma=3.77$~GeV (bottom), for  $^{232}{\rm Th}$, $^{238}{\rm
U}$, and $^{237}{\rm Np}$ (left to right). The same number
($10^5$) of residual nuclei is shown for each case. }
\label{fig:ecn_z377}
\end{centering}
\end{figure}

In summary, as can be seen from Fig.~\ref{fig:0}, when the photon
energy increases from the GDR region to several GeV, the nature of
the photoabsorption process evolves from the excitation of
collective nuclear degrees of freedom to the excitation of a
single nucleon inside the nucleus. In the latter case, up to 95\%
of the photon energy is released in the form of fast particles
leaving the nucleus. Nevertheless, the remaining energy deposited
in the compound nucleus is sufficient for evaporating many
neutrons and thus leading to fission, since neutron evaporation
increases the fissility parameter $Z^2/A$. However,
Fig.~\ref{fig:0} shows only the average values; the entire
distribution of excitation energy $E^\star$ is shown in
Fig.~\ref{fig:ecn_z377}. In high-energy photoabsorption the
$E^\star$ distribution is very wide, since some of the reaction
channels lead to strong heating of the nucleus by a multipion
system.

%% Fig. 2
%%

As shown in Fig.~\ref{fig:0}, the excitation energies of residual
nuclei produced in photoabsorption on lead, thorium, and neptunium
are very close to each other, despite some differences in the
masses of the target nuclei. Therefore, one can expect that the
dramatic differences in fission probabilities in photoabsorption
on such target nuclei are due to differences in fissility
parameters for these ensembles of residual nuclei, rather than in
their excitation energies. Figure~\ref{fig:ecn_z377} confirms this
expectation. In Fig.~\ref{fig:ecn_z377}, the values of $E^\star$
and the liquid-drop-model fissility parameter $Z_{RN}^2/A_{RN}$
are given for the whole ensemble of excited residual nuclei
created after the cascade stage of photoabsorption at $E_\gamma$
values of 68~MeV and 3.77~GeV for  $^{232}{\rm Th}$, $^{238}{\rm
U}$, and $^{237}{\rm Np}$.

As one can see in Fig.~\ref{fig:ecn_z377}, at $E_\gamma=3.77$~GeV,
residual nuclei are created with very broad distributions in
excitation energy, $0<E^\star<300$~MeV, and fissility parameter,
$26<Z_{RN}^2/A_{RN}<38$. At higher photon energies, these broad
distributions are determined mainly by the variations of the
number of pions and nucleons participating in the cascade flow.
Such variations are due to the wide variety of open channels in
the primary $\gamma N$ interaction and in the secondary $\pi N$
and $NN$ interactions with intranuclear nucleons. Taken alone, the
variation in the number of cascade nucleons due to different
impact parameters in a photonuclear interaction cannot provide
such broad distributions. Indeed, the impact-parameter
distributions of photonuclear interaction events are similar for
$E_\gamma=68$~MeV and 3.77~GeV; but in the former case, the
distributions in $Z_{RN}^2/A_{RN}$ are quite narrow. This is
explained by the fact that only a limited number of channels are
open at $E_\gamma=68$~MeV, namely $\gamma+(pn)\rightarrow p+n$ and
channels of the $NN$ interaction below the pion-photoproduction
thresholds.

\begin{figure}[h]
\begin{centering}
{\includegraphics[width=0.8\columnwidth]{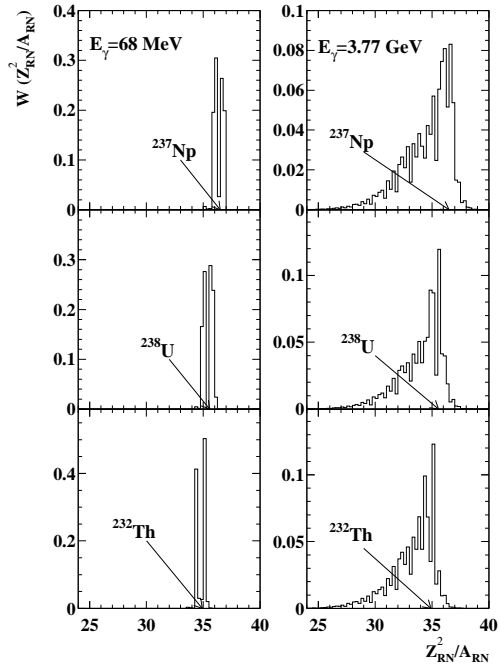}}
\caption{Probability distributions of $Z^2_{RN}/A_{RN}$ for
residual nuclei created after the fast stage of the
photoabsorption process, for $E_\gamma=68$~MeV (left) and for
$E_\gamma=3.77$~GeV (right), for $^{237}{\rm Np}$, $^{238}{\rm
U}$, and $^{232}{\rm Th}$, top to bottom. The values of $Z^2/A$
for the target nuclei are shown by the arrows. }
\label{fig:fiss_par377}
\end{centering}
\end{figure}

\begin{figure}[ht]
\begin{centering}
{\includegraphics[width=0.98\columnwidth]{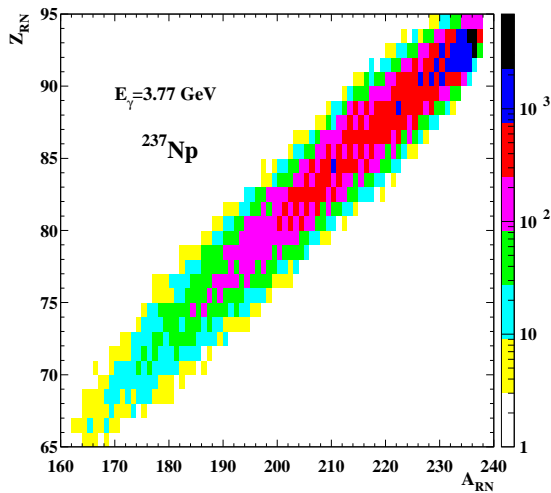}}
\caption{Distribution of masses and charges of
excited residual nuclei created after the
fast stage of the photoabsorption process for $E_\gamma=3.77$~GeV in
 $^{237}{\rm Np}$.
}
\label{fig:a_z377}
\end{centering}
\end{figure}

At higher photon energy, $E_\gamma\geq 1$~GeV, the photoabsorption
process reveals its complexity and the mean-compound-nucleus
approximation used in Ref.~\cite{Arruda-Neto} for
fission-probability calculations breaks down. This is demonstrated
in Fig.~\ref{fig:fiss_par377}, where the fissility-parameter
distributions of the residual nuclei are shown for
$E_\gamma=68$~MeV and 3.77~GeV. Strictly speaking, the
approximation used in Ref.~\cite{Arruda-Neto} is not valid even at
$E_\gamma=68$~MeV, since basically two types of photonuclear
reactions occur: preferential removal of protons, which decreases
$Z_{RN}^2/A_{RN}$, and preferential removal of neutrons, which
increases of $Z_{RN}^2/A_{RN}$. Therefore, the ensemble of
residual nuclei created at low energies, particularly below the
pion-production thresholds, can be represented by a pair of
``characteristic'' compound nuclei. The first group of nuclei,
whose values of $Z_{RN}^2/A_{RN}$ are less than that for the
target nucleus, is created mainly via $(\gamma,p)$, $(\gamma,pn)$,
and $(\gamma,p2n)$ reactions, while the second group is created
via $(\gamma,n)$, $(\gamma,2n)$, and $(\gamma,3n)$ reactions.
However, as shown in Fig.~\ref{fig:fiss_par377}, this
approximation for the ensemble of residual nuclei becomes invalid
in the GeV region, where a very wide set of residual nuclei is
created following the fast stage of photoabsorption. An example of
such a distribution is shown in Fig.~\ref{fig:a_z377} for
photoabsorption in $^{237}{\rm Np}$. In the course of the cascade
process, the target nucleus loses up to $\sim 30$ units in charge
and up to $\sim 70$ nucleons. Highly excited preactinide nuclei,
like Pb or Au, can be created in photoabsorption reactions at high
photon energies. Poorly explored nuclei with $A$ and $Z$ located
between Po and Th are also represented in Fig.~\ref{fig:a_z377}.
In our calculations, we take into account the whole ensemble of
excited compound nuclei without replacing it by a
``characteristic'' average compound nucleus.

%% Fig. 3
%%
%%

%% Fig. 4
%%
%%%

\subsection{Decay of the Excited Compound Nucleus}

As soon as statistical equilibrium is reached in the residual
nucleus, the statistical approach for nucleon and light-particle
evaporation and nuclear fission is an appropriate scheme for
calculation of the relative probabilities of different decay modes
of the compound nucleus. Such statistical decay of the compound
nucleus is the slow stage of the photonuclear reaction, having a
characteristic time $\tau_{evf}\gg \tau_o$. Here we use the term
``residual nucleus'' to mean the nuclear residue formed at the end
of the intranuclear cascade, while the term ``compound nucleus''
is used to describe a residual nucleus after the establishment of
thermal equilibrium.

In the domain of modest excitation energies, which are typical for
heavy nuclei, with $E^\star/A_{CN}\leq 2-3$~MeV, one can include
the competition between evaporation and fission only; the onset of
multifragmentation is located well above this range at
$E^\star/A_{CN}> 4$~MeV~\cite{JPB}. As can be seen in
Fig.~\ref{fig:ecn_z377}, there are no residual nuclei with
$E^\star/A_{RN}$ exceeding $\sim 1.3$~MeV, even at $E_\gamma=3.77$
GeV. Therefore, we do not expect any contribution of the
multifragmentation process to photoabsorption on heavy nuclei.

According to the standard Weisskopf evaporation
scheme~\cite{Weisskopf}, the partial width $\Gamma_j$ for the
evaporation of a particle $j=n,p,{\rm ^2H},{\rm ^3H},{\rm ^3He}$,
or ${\rm ^4He}$ is given by
\begin{equation}
\Gamma _j = {{(2s_j + 1) \mu_j} \over {\pi ^2 \rho_{CN} (E^\star)}}
\int \limits_{V_j}^{E^\star - B_j} \sigma_{inv}^j (E)
\rho_j (E^\star - B_j - E)
E dE \mbox{ ,}
\label{GE}
\end{equation}
where $s_j$, $\mu_j$, $V_j$, and $B_j$ are the spin, reduced mass,
Coulomb barrier, and binding energy of the particle $j$,
respectively. $\sigma_{inv}^j (E)$ is the cross section for the
inverse reaction of the capture of the particle $i$ to create the
compound nucleus. $\rho_{CN}$ and $\rho_j$ are the nuclear level
densities for the initial and final (after the emission of the
particle $j$) nuclei, respectively. Natural units, with
$\hbar=c=1$, are used in this paper.

The Bohr-Wheeler statistical approach~\cite{Bohr-Wheeler} is used
to calculate the fission width of the excited compound nucleus.
This width is proportional to the nuclear level density $\rho_f$
at the fission saddle point:
\begin{equation}
\Gamma _f = {1 \over {2 \pi \rho_{CN} (E^\star)}}
\int \limits_0^{E^\star - B_f} \rho_f (E^\star - B_f - E) dE \mbox{ ,}\\
\label{GF}
\end{equation}
where $B_f$ is the fission barrier height. Masses and kinetic
energies of fission products are calculated based on corresponding
approximations~\cite{Adeev} to experimental data which describe
the transition from the asymmetric fission mode to the symmetric
one, along with other features of the fission process. Since the
masses of fission fragments were not measured in the experiments
reported in Refs.~\cite{Sanabria,SanabriaPhd,Cetina,Cetina2002},
we do not consider such distributions in this work.

The decay of the excited compound nucleus is simulated using the
Monte-Carlo method. The competition between the various decay
channels at each step of the evaporation chain is determined by
the relation between the partial widths for particle evaporation
and fission, Eqs.~(\ref{GE}) and (\ref{GF}), respectively.
Finally, in order to calculate the fission probability $W_f$, the
total number of fission events in a computer run is counted and
divided by the total number of simulated photoabsorption events.
Evaporation from excited fission fragments is also taken into
account and was found to be negligible.

In the present paper, we also take into account the microscopic
effects of nuclear structure in the nuclear-mass and level-density
formulas, according to Refs.~\cite{Ignatyuk75,Ignatyuk79,ILJ-MEB}.
Such effects reveal themselves as a noticeable difference, up to
$\sim 10-15$~MeV for heavy closed-shell nuclei, between the values
of the measured nuclear ground-state masses $M_{exp}(Z,N)$ and
those predicted by the macroscopic liquid-drop model
$M_{LD}(Z,N)$~\cite{MS}. Moreover, this difference in mass $\delta
W_{gs}=M_{exp}(Z,N)-M_{LD}(Z,N)$, the so-called shell correction,
and the level-density parameter $a$ (used in Eq.~(\ref{Fermi})
below) are strongly correlated. For closed-shell nuclei, the
actual values of the level-density parameter are substantially
lower than the average values of $A/8-A/10$~MeV$^{-1}$, and these
values depend strongly on the excitation energy. Proper accounting
for these effects, as well as for pairing effects, is important
mainly at low excitations, for $E^\star \sim 10$~MeV. Although
these shell effects are very pronounced at low excitation
energies, they disappear for $E^\star\geq
30$~MeV~\cite{ILJ-MEB,Gaimard}.  Several phenomenological
approximations of the level-density parameter were proposed in
order to account for such behavior. Our calculations are based
mainly on the results of Ref.~\cite{ILJ-MEB}, where the data on
the level densities, decay widths, and lifetimes of excited nuclei
with $2<E^\star<20$~MeV have been analyzed in the framework of the
statistical model.

We use the Fermi-gas expression for the nuclear level density at excitation energy $E^\star$:
\begin{equation}
\rho(E^\star)={\sqrt{\pi}\over{12E^{\star 5/4}a^{1/4}}} \cdot
\exp\{2\sqrt{aE^\star} \} \mbox{ ,}
\label{Fermi}
\end{equation}
where $a=\pi^2g_F/6$ is the nuclear level-density parameter,
which is proportional to the density of single-particle
states $g_F$ at the Fermi surface.

Pairing-energy effects are accounted for by the substitution $E^\star\rightarrow E^\star-\Delta$, where the pairing
energy $\Delta$ is given by:
\begin{equation}
\Delta= \chi\cdot 11/\sqrt{A} {\rm\mbox{  }MeV}
\end{equation}
\noindent with $\chi=0$, 1, or 2 for odd-odd, odd-even, or even-even nuclei,
respectively.

The nuclear level-density parameter is a function of $Z$, $N$, and $E^\star$ of the corresponding
nucleus~\cite{Ignatyuk75,Ignatyuk79,ILJ-MEB}:
\begin{equation}
a(Z,N,E^\star) = \tilde a(A) \left\{ 1 + \delta W_{gs}(Z,N)
{{f(E^\star - \Delta)} \over {E^\star - \Delta}} \right\}\mbox{, }
\end{equation}
\noindent where
\begin{equation}
f(E)=1-\exp(-\gamma E)\mbox{; }
\end{equation}
\begin{equation}
\tilde a(A) = \alpha A+\beta A^{2/3}B_s
\label{atild}
\end{equation}
is the asymptotic Fermi-gas value at high excitation energies, and
$\delta W_{gs}(Z,N)$ is the shell correction in the nuclear mass
formula. The coefficients $\alpha$ and $\beta$ correspond to the
volume and surface components, which, along with $\gamma$, are
taken to be phenomenological constants. We use the values
$\alpha=0.114$, $\beta=0.098$, and $\gamma=0.051$ (all in
MeV$^{-1}$), corresponding to the first set of systematics given
in Table~3 of Ref.~\cite{ILJ-MEB}, which includes the shell
corrections of Ref.~\cite{MS}. $B_s$ is the surface area of the
nucleus in units of the surface for a sphere of equal volume.
$B_s\approx 1$ for nearly spherical nuclei with small deformation.
Since the surface is systematically larger at the saddle point,
$B_s>1$ there; in fact $B_s\approx 2^{1/3}\approx 1.26$ for the
configuration corresponding to the splitting of the compound
nucleus into two equal spherical fragments. Such a case nearly
corresponds to the fission of low-fissility nuclei, while for
highly fissile nuclei the shape at the saddle point is closer to a
sphere and $B_s\approx 1$.

In Ref.~\cite{Martins}, for example, the level-density parameter
$a_f$ at the fission saddle point (in the transition state) is
calculated using an analogous parameter for the neutron emission
channel $a_n$. In this way, a constant ratio $r=a_f/a_n$ is
assumed and used as a fitting parameter of the model. As a
consequence of this assumption, the shell-effect influence on the
level density for the neutron-emission channel is transferred to
the level density at the saddle point. However, contrary to
Ref.~\cite{Martins}, we expect that the shell corrections at the
saddle point with large deformation have no relation to those at
the ground state with relatively small equilibrium
deformation~\cite{Cherepanov,Ignatyuk75,Ignatyuk79}. We conclude
that the shell corrections should be much smaller at the saddle
point and therefore we can use the asymptotic value
$\tilde{a}_f(A)$ instead of an energy-dependent
$a_f(Z,N,E^\star)$. The value of $\tilde{a}_f(A)$ is assumed to be
proportional to the asymptotic one for the neutron-emission
channel $\tilde{a}_n(A)$.

\begin{figure}[ht]
\begin{centering}
{\includegraphics[width=0.95\columnwidth]{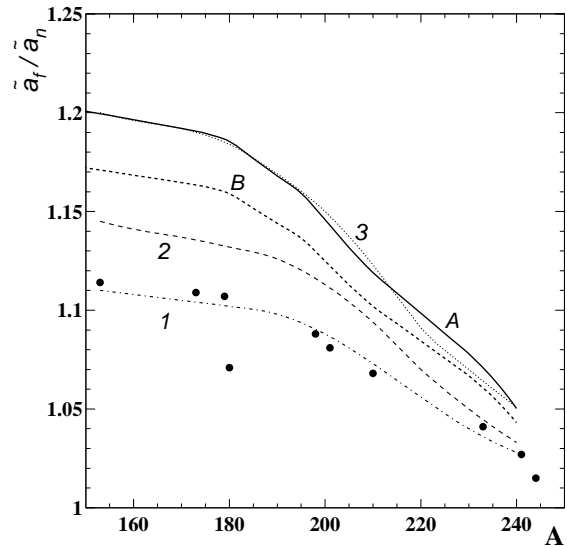}} \caption{The
closed circles are the $\tilde{a}_f/\tilde{a}_n$ ratios calculated
in Ref.~\cite{Reisdorf} based on the liquid-drop model and the
dot-dashed curve marked ``1'' shows the result of our
interpolation of these discrete ratios. The two sets, ${\cal A}$
and ${\cal B}$, of $\tilde{a}_f/\tilde{a}_n$ values used in our
calculations are represented by the solid and short-dashed and
dotted curves, respectively. For comparison, calculational results
from Refs.~\cite{Toke,Treiner} are shown by the long-dashed and
dotted curves, marked as ``2'' and ``3,'' respectively. }
\label{fig:afan}
\end{centering}
\end{figure}

The ratios $\tilde{a}_f/\tilde{a}_n$ for some nuclei were
calculated in Ref.~\cite{Reisdorf} based on the liquid-drop
model~\cite{MS74} and using an expression similar to
Eq.~(\ref{atild}), but with an additional curvature term. As shown
in Fig.~\ref{fig:afan}, the ratios obtained from this procedure
are very close to unity for the actinides ($\sim 1.02-1.04$) and
are higher for the preactinides ($\sim 1.07-1.12$). Since the
value for $\tilde{a}_f/\tilde{a}_n$ used in our calculations needs
to be valid for a much wider range of nuclides, we tried to
interpolate between the values tabulated in Ref.~\cite{Reisdorf}
to obtain the values for other nuclei as shown, for example, in
Fig.~\ref{fig:a_z377}. Such a procedure deals with the average
values of $\tilde{a}_f/\tilde{a}_n$ and neglects possible rapid
changes of the ratio for individual isotopes. However, this is the
only method to estimate $\tilde{a}_f/\tilde{a}_n$ for the region
of nuclei between Po and Th, where experimental information on the
nuclear level densities is lacking.

%% Fig. 5
%%
%%%

Using this interpolation procedure, we find that the photofission
cross sections for lead are underestimated by theory by a large
factor. Therefore, to improve the agreement with the experimental
data~\cite{SanabriaPhd,Cetina2002}, we increased the difference
between the values of Ref.~\cite{Reisdorf} and unity by a factor
of 1.7 (variant $\cal A$) or 1.5 (variant $\cal B$). The resulting
curves are shown in Fig.~\ref{fig:afan}. As one can see, such
modification is not too far outside the range of
$\tilde{a}_f/\tilde{a}_n$ values which were obtained in
Refs.~\cite{Reisdorf,Toke,Treiner}. Moreover, the curve of variant
$\cal A$ independently obtained from the analysis of the
photofission data on $^{\rm nat}{\rm
Pb}$~\cite{SanabriaPhd,Cetina2002}, agrees very well indeed with
the values of Ref.~\cite{Treiner}, the most recent and
comprehensive of these studies of nuclear level densities, denoted
by curve ``3'' in Fig.~\ref{fig:afan}.

Following Refs.~\cite{Ignatyuk75,Ignatyuk79,ILJ-MEB}, the fission
barriers are computed from the macroscopic liquid-drop component
$B_{LD}(Z,N)$ and the shell correction at the nuclear ground state
$\delta W_{gs}(Z,N)$~\cite{MS}:
\begin{equation}
B_f(Z,N,E^\star)=B_{LD}(Z,N)-\delta W_{gs}(Z,N).
\end{equation}
In this expression, the shell correction a the saddle point is
neglected.

The validity of the above-mentioned assumptions (except for our
interpolation procedure for $\tilde{a}_f/\tilde{a}_n$) was
confirmed in previous studies of nuclear fission induced by
various projectiles. For example, the dependence of $W_f$ on the
target-nucleus mass for reactions with stopped $\pi^-$ mesons,
photons, protons, and $\alpha$-particles with kinetic energies
below 1~GeV was considered in Ref.~\cite{Cherepanov} within the
framework of the intranuclear cascade, fission, and evaporation
models. The calculated values were compared with the experimental
data available at that time. The influences on fission probability
of shell effects, pre-equilibrium emission from the residual
nucleus, and the parameters of the liquid-drop model were studied.
Later, the $W_f$ values were deduced  within this model for heavy
nuclei ($Z^2/A>30$) in reactions with these projectiles and with
pions in flight~\cite{Iljinov-Mebel}.

\subsection{Results for electromagnetically-induced heavy-ion fission}

As shown above, when energetic photons are absorbed by heavy nuclei,
a wide range of excited residual nuclei is created, having masses and
charges far from the stability line. Beams of relativistic
radioactive nuclei provide new access to fission studies of such
nuclei. For example, in Ref.~\cite{Heinz}, fission of 58 secondary
projectiles ($^{231-234}$U, $^{226-231}$Pa, $^{221-229}$Th,
$^{215-226}$Ac, $^{211-223}$Ra, $^{208-212,217,218}$Fr,
$^{205-209}$Rn, and $^{205,206}$At) produced by fragmentation of
1~GeV/nucleon $^{238}$U nuclei has been studied. The total fission
cross section in secondary collisions of these nuclei with a lead
target at energy 420~MeV/nucleon was found to consist of two
comparable parts, due to hadronic and electromagnetic
interactions.

Electromagnetically-induced fission results from absorption of
virtual photons. According to the RELDIS model, which is based on
the Weizs\"{a}cker-Williams method of equivalent quanta, such
nuclei are typified by low excitations, $\langle E^\star
\rangle\sim 12-13$ MeV, due to the photons in the GDR region with
energies below 20 MeV. The recent data~\cite{Heinz} make possible
a crucial test of the model, since  $E^\star\sim B_f$ in such
processes. This makes the model results very sensitive to the
fission barrier shape and height, as well as to the level-density
parameterization $\rho(E^\star)$ used in the calculations.

The experimental fission probabilities $w^{ED}_f$ for each of the
ions were derived by dividing the measured fission cross sections
of Ref.~\cite{Heinz} by the calculated total
electromagnetic-dissociation cross sections $\sigma^{ED}$. The
calculations were performed by accounting for absorption of either
one or two photons in each collision event; see
Refs.~\cite{Pshenichnov,Pshenichnov2,Pshenichnov2001} for details.
Double-photon absorption processes were taken into account by
applying the harmonic-oscillator ansatz in conjunction with the
folding model~\cite{Llope:vp}.

Since experimental data on the total photoabsorption cross
sections for unstable nuclei are not available, the cross sections
for nearby stable nuclei were used in calculations of
$\sigma^{ED}$. Approximations of the total photoabsorption cross
sections obtained in Ref.~\cite{Berman-Fultz} were used. Such a
substitution generally leads to 2-3\% error in $\sigma^{ED}$
according to the estimates based on the GDR sum
rule~\cite{Berman-Fultz,ADNDT}.

\begin{figure}[ht]
\begin{centering}
{\includegraphics[width=0.85\columnwidth]{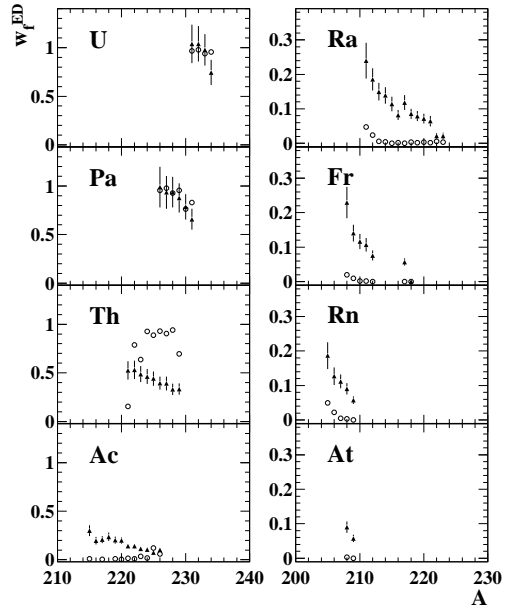}}
\caption{Fission probabilities in electromagnetic dissociation of
heavy ions at 420 MeV/nucleon on a lead target.
Experimental fission probabilities extracted from the
fission cross section of Ref.~\cite{Heinz} are shown by the full
triangles. Calculational results are given by the open circles.}
\label{fig:wf_emf}
\end{centering}
\end{figure}

Experimental and calculated values of $w^{ED}_f$ are shown in
Fig.~\ref{fig:wf_emf}. Very good agreement is found for the most
fissile nuclei $^{231-234}$U and $^{226-231}$Pa. The description
of the data for Th nuclei is poor. The calculated values for Ac,
Ra, Fr, Rn, and At are much lower than the experimental data, but
good agreement is obtained for the values of $w^{ED}_f$ for those
isotopes which are closest to the stable isotopes of Ac, Ra, and
Fr. This is explained by the fact that the input calculational
parameters were adjusted to describe the fission of nuclei close
to the stability line. The calculations presented in
Ref.~\cite{Heinz} for Ra nuclei also underestimate the
electromagnetic fission cross sections. At the present time, the
description of the fission of radioactive nuclei seems to present
a common difficulty to all fission models. Further work on our
fission model should be aimed at a better description of the
fission of unstable nuclei.
%% Fig.6

\begin{figure}[ht]
\begin{centering}
{\includegraphics[width=0.85\columnwidth]{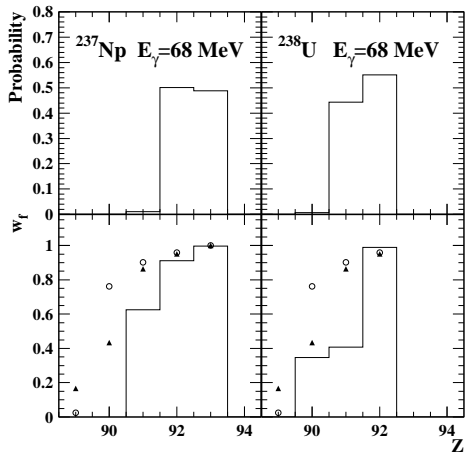}}
\caption{Top:
charge distributions of residual nuclei created in photoabsorption
on $^{237}$Np and $^{238}$U at $E_\gamma=68$ MeV. Bottom: fission
probabilities for such nuclei. Calculated fission probabilities in
photoabsorption at $E_\gamma=68$ MeV are shown by histograms.
Calculated and measured fission probabilities for heavy ions at 420 MeV/nucleon
are shown by open circles and filled triangles, respectively.}
\label{fig:wf_z}
\end{centering}
\end{figure}

Using single-humped fission barriers instead of double-humped
barriers and neglecting the collective effects in the nuclear
level-density formula does not lead to disagreement with the data
for U and Pa nuclei. Below we estimate how the failure to describe
the fission of other radioactive nuclei may affect the total
calculated fission probability $W_f$ in phototoabsorption. The 58
radioactive nuclei listed are only a small subset of all of the
residual nuclei created in photoabsorption of GeV photons. As a
rule, such residual nuclei are much more highly excited. However,
photoabsorption at lower energies, such as at $E_\gamma=68$ MeV,
creates nuclei with $\langle E^\star \rangle$ of only about 20
MeV, much closer to nuclei with $\langle E^\star \rangle\sim
12-13$ MeV, which is the case for electromagnetically induced
heavy-ion fission. This makes useful the comparison presented in
Fig.~\ref{fig:wf_z}.
%% Fig.7

The charge distributions of residual nuclei created in
photoabsorption on $^{237}$Np and $^{238}$U at $E_\gamma=68$ MeV
are shown in Fig.~\ref{fig:wf_z}, together with the calculated
fission probabilities for residual elements. Excited Np, U, Pa,
and Th nuclei are created  in photoabsorption at this energy. The
calculated and measured values of $w^{ED}_f$ for such nuclei in
electromagnetically induced fission are shown in the bottom part
of the same plot. These values are obtained as average values of
the points for individual nuclides, and they are different from
the values of $w_f$ for photofission, due to the difference in
$E^\star$.  As shown, fission probability in electromagnetically
induced fission of U and Pa is described at the 5\% level of
accuracy. The agreement for Th is poor, but this element is less
abundant in the ensembles of residual nuclei; the estimate of the
total fission probability $W_f$ for the whole ensemble is quite
reliable.

The reliability of the fission model can be quantitatively
assessed by calculating the total fission probability for the
ensembles presented in Fig.~\ref{fig:wf_z}. In the following two
hypothetical examples the experimental and theoretical values for
$w^{ED}_f$ were used, but in both cases the residual-element
abundances were taken for photoabsorption at $E_\gamma=68$ MeV. As
found for the $\gamma + $Np case, $W_f=0.98$ for the calculated
$w^{ED}_f$, compared with $W_f=0.97$ for the value extracted from
experiment. For the $\gamma +$U case, $W_f=0.93$ and $W_f=0.91$,
respectively, for the two sets of input $w^{ED}_f$ values.
Therefore, the failure to describe the fissility of Th leads to an
error in $W_f$ of only $\sim3$\%. In this way, the reliability of
the fission model is demonstrated  for the most fissile and most
probable residual nuclei created in photoabsorption on $^{237}$Np
and $^{238}$U at low energies.

\section{Results and Discussion}\label{Res_Disc}

\subsection{Fission Probabilities for $^{\bf 237}{\bf Np}$,
$^{\bf 233,235,238}{\bf U}$, $^{\bf 232}{\bf Th}$, and
$^{\bf nat}{\bf Pb}$}

\begin{figure}[ht]
\begin{centering}
{\includegraphics[width=0.85\columnwidth]{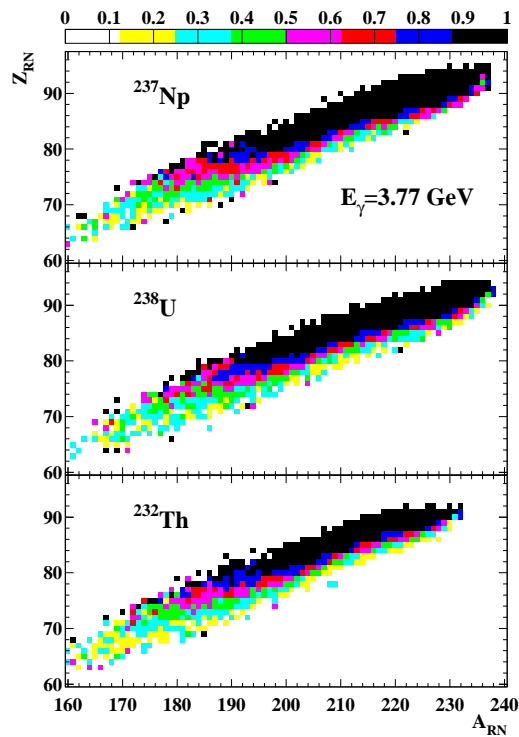}}
\caption{Fission probabilities $w_f$ of each of the residual
nuclei created after the fast stage of the photoabsorption process
for $E_\gamma=3.77$~GeV on $^{237}{\rm Np}$, $^{238}{\rm U}$, and
$^{232}{\rm Th}$ (top, middle, and bottom panels, respectively). }
\label{fig:pfa_z377}
\end{centering}
\end{figure}

As a rule, the liquid-drop model predicts proton-rich nuclei to
have higher probability to undergo fission~\cite{MS,MS74}. The
probability $w_f$ for a residual nucleus with given mass $A_{RN}$
and charge $Z_{RN}$ to undergo fission during the last stage of
the reaction is shown in Fig.~\ref{fig:pfa_z377} for
$E_\gamma=3.77$~GeV for $^{237}{\rm Np}$, $^{238}{\rm U}$, and
$^{232}{\rm Th}$.

%% Fig. 8
%%

\begin{figure}[ht]
\begin{centering}
{\includegraphics[width=1.0\columnwidth]{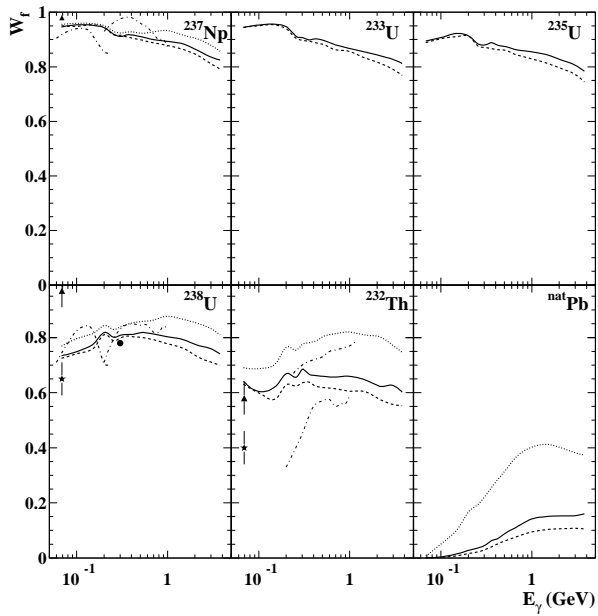}} \caption{Fission
probabilities $W_f$ for photoabsorption in $^{237}{\rm Np}$,
$^{233}{\rm U}$, $^{235}{\rm U}$, $^{238}{\rm U}$, $^{232}{\rm
Th}$, and $^{\rm nat}{\rm Pb}$. Our calculated results for
variants ${\cal A}$ and ${\cal B}$ are given by the solid and
dashed lines, respectively; variant ${\cal A}$ without
pre-equilibrium emission is represented by the dotted lines.
Earlier calculational results for $^{237}{\rm Np}$, $^{238}{\rm
U}$, and $^{232}{\rm Th}$~\cite{Deppman:2002} are given by thin
dot-dashed lines; for $^{237}{\rm Np}$, $^{238}{\rm
U}$~\cite{Iljinov-Nedorezov}, and for $^{232}{\rm
Th}$~\cite{Bianchi93} are given by thick dot-dashed lines. The
circle represents the calculational result of
Ref.~\cite{Lucherini}. Values inferred in Ref.~\cite{Martins} from
the old experimental data of Refs.~\cite{Lep-fiss,Martins} are
given by the triangles and stars, respectively. } \label{fig:pf}
\end{centering}
\end{figure}

The probability $w_f$ is defined for each of the nuclides created
after the INC stage of photoabsorption, while the fission
probability $W_f$ refers to the appropriately weighted average
value calculated over the whole ensemble of residual nuclei. For
example, $w_f$ turns out to be below 0.2 for some regions of
$A_{RN}$ and $Z_{RN}$ far from the $A$ and $Z$ of the initial
target nucleus. However, these regions do not contribute much to
the resulting $W_f$, since the probability to create a nucleus
with such $A_{RN}$ and $Z_{RN}$ is low, as can be seen in
Figs.~\ref{fig:a_z377} and~\ref{fig:pfa_z377}. As a result, the
$W_f$ values, shown in Fig.~\ref{fig:pf} as functions of
$E_\gamma$ for $^{237}{\rm Np}$, $^{238}{\rm U}$, and $^{232}{\rm
Th}$, are generally above 0.6 despite the fact that some of the
residual nuclei have a low probability $w_f$ to undergo fission.
As is also shown in Fig.~\ref{fig:pf}, this is true for
$^{233}{\rm U}$ and $^{235}{\rm U}$ as well, for which $0.8 \leq
W_f\leq 0.95$, almost as large as the corresponding values for
$^{237}{\rm Np}$.

%% Fig. 9
%%

Our calculated values for $W_f$ for $^{\rm nat}{\rm Pb}$ are much
lower, $\sim 0.10-0.15$, as can be seen in Fig.~\ref{fig:pf}. The
values for $W_f$ are first obtained for each of the most abundant
lead isotopes, $^{206}{\rm Pb}$ (24.1\%), $^{207}{\rm Pb}$
(22.1\%), and $^{208}{\rm Pb}$ (52.4\%), and then are averaged
with appropriate weights to obtain the fission probability for
$^{\rm nat}{\rm Pb}$.

As can be seen in Fig.~\ref{fig:pf}, two extreme cases are
represented by photoabsorption on the highly fissile actinides
$^{237}{\rm Np}$, $^{233}{\rm U}$, and $^{235}{\rm U}$ and on the
pre-actinide $^{\rm nat}{\rm Pb}$. At $E_\gamma=68$~MeV, fission
clearly dominates for the residual nuclei created after
photoabsorption on $^{237}{\rm Np}$, $^{233}{\rm U}$, and
$^{235}{\rm U}$. However, the fission contribution gradually
decreases with increasing $E_\gamma$, since, on average, a wider
distribution of residual nuclei created at higher $E_\gamma$ has
lower fission probability and nucleon evaporation becomes more
important. A completely different tendency is found for the
photofission of lead, which has a clear threshold behavior. Due to
high fission barriers for the residual nuclei created after
photoabsorption on $^{\rm nat}{\rm Pb}$, the fission process is
suppressed below $E_\gamma\sim 100$ MeV and only evaporation from
excited nuclei can take place there. Above $E_\gamma\sim 100$ MeV,
the fission probability $W_f$ gradually increases to $W_f\sim
0.1$, but this still represents only a small part of the
deexcitation process. An intermediate tendency is found in
photoabsorption on $^{238}{\rm U}$ and $^{232}{\rm Th}$. For these
cases, $W_f$ has a very broad maximum at $E_\gamma\sim 0.5-1$~GeV
and decreases gradually with $E_\gamma$ above $\sim 1$~GeV.

Different mechanisms of photoabsorption play a role at different
photon energies. As a result, a variation in the trend of the
average excitation energy versus $E_\gamma$ is found for
180--250~MeV, as seen in Fig.~\ref{fig:0}. The variations of
$W_f(E_\gamma)$ which are seen in Fig.~\ref{fig:pf} have the same
origin. The probability for a pion produced in the $\gamma
N\rightarrow\pi N$ process to be absorbed in the nuclear medium
increases rapidly with $E_\gamma$ as one approaches the region of
the $\Delta_{33}(1232)$ resonance, $E_\gamma\sim 200-500$~MeV. If
this pion is re-absorbed in the nucleus, the excitation energy
increases and, therefore, so does the fission probability.

If the pre-equilibrium emission process after the cascade stage of
photonuclear reaction is neglected in the calculations, the decay
of the excited compound nucleus takes place with higher excitation
energy $E^\star$ and fissility parameter $Z^2/A$. This generally
leads to higher values of $W_f$, as shown in Fig.~\ref{fig:pf} by
dotted lines for $^{237}{\rm Np}$, $^{238}{\rm U}$, $^{232}{\rm
Th}$, and $^{\rm nat}{\rm Pb}$.

The changes in $W_f$ due to the variation of
$\tilde{a}_f/\tilde{a}_n$ (${\cal A}$ and ${\cal B}$) are of the
order of 5\% for $^{237}{\rm Np}$,  $^{233}{\rm U}$, $^{235}{\rm
U}$, and $^{238}{\rm U}$ and 10\% for $^{232}{\rm Th}$, as also
can be seen in Fig.~\ref{fig:pf}. This important observation of
the relative stability of the calculated $W_f$ is true for
actinide nuclei only. This makes it possible to infer the total
photoabsorption cross section from the photofission
data~\cite{Sanabria,Cetina,Cetina2002} and the calculated values
for $W_f$.

Fission probabilities estimated in Refs.~\cite{Martins,Lep-fiss}
from previous experimental data on absolute photofission cross
sections are given in Fig.~\ref{fig:pf} for comparison, although
they are in poor agreement with each other. The calculated $W_f$
value from Ref.~\cite{Lucherini} is in good agreement with our
results for $^{238}{\rm U}$.

Compared with the approach of
Refs.~\cite{ILJ-MEB,Iljinov-Nedorezov}, several simplifications
were adopted in our fission calculations. For example, we do not
use double-humped fission barriers for actinides, nor do we take
into account any collective effects in the nuclear level-density
formula because (i) information (either experimental or
theoretical) on such effects is not available for the whole range
of residual nuclei, and (ii) at high excitation energies
($E^\star\geq 30$~MeV), which are our interest here, the fission
probability $W_f$ is sensitive mainly to the ratio
$\tilde{a}_f/\tilde{a}_n$. The uncertainty in the choice of
$\tilde{a}_f/\tilde{a}_n$ exceeds the uncertainties resulting from
the selection of other calculational parameters describing the
decay of the excited residual nucleus.

As shown in  Fig.~\ref{fig:pf}, the difference between our
approach and that of Refs.~\cite{ILJ-MEB,Iljinov-Nedorezov} does
not lead to a dramatic difference in $W_f$ for $^{237}{\rm Np}$
and $^{238}{\rm U}$. In the limited energy region $60\leq E_\gamma
\leq 240$ MeV, $W_f\sim 85-95$\% and $\sim 70-85$\%, respectively,
for both approaches. However, for $250 \leq E_\gamma \leq 1200$
MeV,  our fission probabilities  for $^{232}{\rm Th}$ (variants
${\cal A}$ and ${\cal B}$) are noticeably lower than those of
Ref.~\cite{Bianchi93}, where the effect of pre-equilibrium
emission, the trawling effect, and some multiple-pion
photoproduction channels were neglected.

\begin{figure}[ht]
\begin{centering}
{\includegraphics[width=0.7\columnwidth]{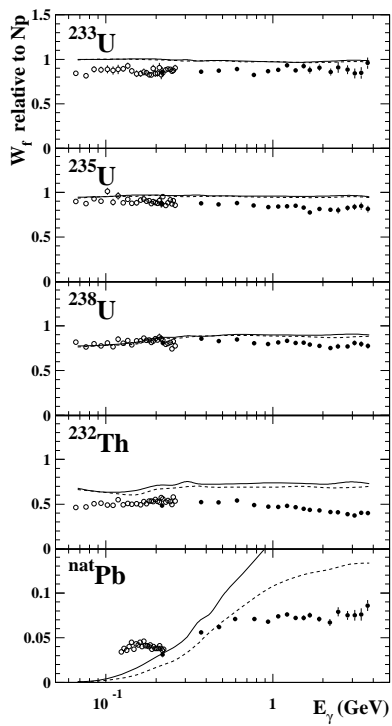}}
\caption{Fission probabilities relative to $^{237}{\rm Np}$ for
photoabsorption in $^{233}{\rm U}$, $^{235}{\rm U}$ $^{238}{\rm
U}$, $^{232}{\rm Th}$, and $^{\rm nat}{\rm Pb}$. Our calculated
results for variants ${\cal A}$ and ${\cal B}$ are given by the
solid and dashed lines, respectively. The open and closed circles
represent the data from SAL~\cite{Sanabria,SanabriaPhd} and
JLab~\cite{Cetina,Cetina2002}, respectively. }
\label{fig:rel_u_pb}
\end{centering}
\end{figure}

Our calculated fission probabilities relative to $^{237}{\rm Np}$
are shown in Fig.~\ref{fig:rel_u_pb} for the uranium isotopes,
thorium, and lead as a function of $E_\gamma$, compared with the
experimental data from SAL~\cite{Sanabria,SanabriaPhd} and
JLab~\cite{Cetina,Cetina2002}. Again, only minor changes are found
due to the choice of $\tilde{a}_f/\tilde{a}_n$ in variants ${\cal
A}$ and ${\cal B}$ for actinides, whose energy dependence of
relative fission probabilities is flat.

%% Fig. 10
%%

A better description of the experimental fission probabilities
relative to $^{237}{\rm Np}$ is found for $^{233}{\rm U}$,
$^{235}{\rm U}$, and $^{238}{\rm U}$. Larger uncertainties in the
calculations exist for $^{232}{\rm Th}$, and especially for $^{\rm
nat}{\rm Pb}$, where the $W_f$ values are farther from unity.

As shown in Figs.~\ref{fig:pf} and~\ref{fig:rel_u_pb}, our
calculated fission probabilities for $^{237}{\rm Np}$ and
$^{238}{\rm U}$ are in agreement within $\sim 10$--20\%  with the
theoretical results of other authors and the estimates based on
earlier experimental data from the literature, which are available
mainly at low energies, $E_\gamma < 300$ MeV. The fission
probabilities for $^{237}{\rm Np}$ and $^{233,235,238}{\rm U}$ are
stable within $\sim 7$\% with respect to the choice of
calculational parameters. At $E_\gamma\geq 100$~MeV, the main
uncertainty in our results for these nuclei is due to the choice
of $\tilde{a}_f/\tilde{a}_n$.

For the case of $^{232}{\rm Th}$, we see a larger discrepancy with
the data than for the other actinides, while the energy dependence
is still reproduced reasonably well. The $^{232}{\rm Th}$ nucleus
deviates from the pattern common to the heavier actinides at lower
energies with regard to fission probability~\cite{Caldwell} and
the number of prompt neutrons emitted per fission~\cite{Caldwell1}
because it has a more spherical shape~\cite{Caldwell}. It may be
that this deviation persists to GeV energies as well.

\subsection{The Total Photoabsorption Cross Section as a Calculational Input}\label{total}

In order to obtain the absolute photofission cross section, the
calculated fission probability must be multiplied by the total
photoabsorption cross section. In the RELDIS code, the values of
the total photoabsorption cross section are taken from
approximations to the existing experimental data. In the GDR
region, the Lorentz-curve fits with parameters from
Refs.~\cite{Berman-Fultz,ADNDT}, corrected according to the
prescription of Ref.~\cite{Berman}, were used for this purpose.
Above the GDR region, where quasideuteron absorption becomes
dominant, the total cross section is taken from the estimate of
Ref.~\cite{Lepretre} (based on the quasideuteron
model~\cite{Levinger}.)

Above the pion-production thresholds, a ``universal'' behavior,
$\sigma_{\gamma A}(E_\gamma)\propto A$, is observed (see
Ref.~\cite{Bianchi,Valeria} for the latest experimental data).
This means that the total photoabsorption cross section per bound
nucleon $\sigma_{\gamma A}(E_\gamma)/A$ has the same magnitude and
the same energy dependence for light, medium, and heavy nuclei (C,
Al, Cu, Sn, and Pb) at least up to $E_\gamma \sim 3$~GeV.
Therefore, having the data for one nucleus, one can calculate the
cross section for other nuclei. However, in this energy region the
universal curve $\sigma_{\gamma A}(E_\gamma)/A$ is very different
from the values extrapolated from the cross sections on free
nucleons, $(Z\sigma_{\gamma p}+N\sigma_{\gamma n})/A$, which are
deduced from proton~\cite{Daresbury,DAPHNE} and deuteron
data~\cite{Daresbury:gn}. For $E_\gamma > 3$~GeV, the universal
behavior breaks down, and, for example, the ratio $\sigma_{\gamma
A}(E_\gamma)/A$ for lead is 20-25\% lower than for
carbon~\cite{Engel-Roesler} due to the nuclear shadowing
effect~\cite{Weise,Brooks}. In order to approximate the total
photonuclear cross sections at $E_\gamma > 3$~GeV, we use recent
results for $\sigma_{\gamma A}(E_\gamma)/A$ obtained with the
Glauber-Gribov approximation within the Generalized Vector
Dominance Model (see Ref.~\cite{Engel-Roesler} and references
therein).

\subsection{Absolute Photofission Cross Sections for $^{\bf 237}{\bf Np}$, $^{\bf 233,235,238}{\bf U}$,
$^{\bf 232}{\bf Th}$, and $^{\bf nat}{\bf Pb}$}

\begin{figure}[ht]
\begin{centering}
{\includegraphics[width=1.0\columnwidth]{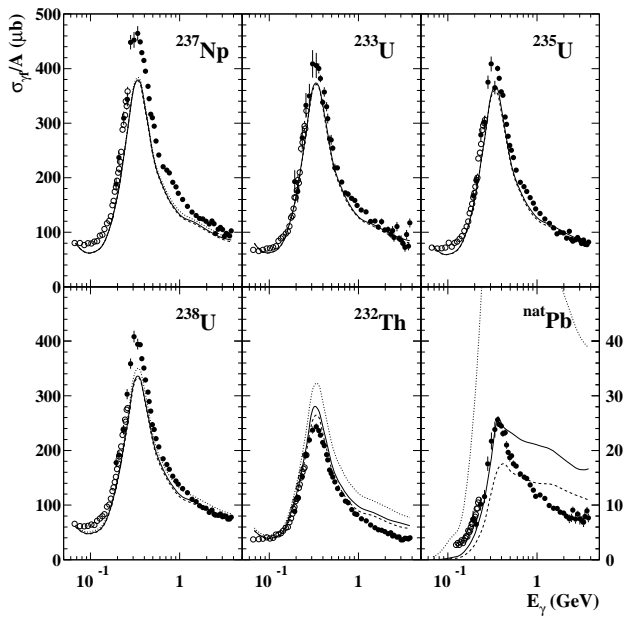}}
\caption{Absolute photofission cross sections per bound nucleon
for $^{237}{\rm Np}$, $^{233,235,238}{\rm U}$, $^{232}{\rm Th}$,
and  $^{\rm nat}{\rm Pb}$. Our calculated results for variants
${\cal A}$ and ${\cal B}$ are given by the solid and dashed lines,
respectively. Variant ${\cal A}$ without pre-equilibrium emission
is represented by the dotted lines. Cross-section values given in
the plot for $^{\rm nat}{\rm Pb}$ (only) are shown on a separate
scale on the right-hand side. The data are from
Refs.~\cite{Sanabria,SanabriaPhd,Cetina,Cetina2002}. }
\label{fig:abs_bw}
\end{centering}
\end{figure}

Our calculated absolute photofission cross sections for
$^{237}{\rm Np}$, $^{233,235,238}{\rm U}$, $^{232}{\rm Th}$, and
$^{\rm nat}{\rm Pb}$ are shown in Fig.~\ref{fig:abs_bw} and
compared with the experimental data of
Refs.~\cite{Sanabria,SanabriaPhd,Cetina,Cetina2002}. For
$^{237}{\rm Np}$ and $^{233,235,238}{\rm U}$, variants ${\cal A}$
and ${\cal B}$ give consistent results. Except for $^{232}{\rm
Th}$, neglecting pre-equilibrium emission after the INC stage does
not lead to a large change in the calculated photofission cross
section; a small increase in the calculated cross section is
obtained only for $E_\gamma\agt 1$~GeV. When normalized to the
``Universal Curve'' in this way, our calculated absolute cross
sections agree within $\sim 7$\% with the data for $^{233}{\rm U}$
and $^{235}{\rm U}$, but are underestimated for $^{237}{\rm Np}$
and $^{238}{\rm U}$ and overestimated for $^{232}{\rm Th}$,
especially in the $\Delta_{33}(1232)$-resonance region, although
qualitative agreement for the shape of the cross sections is
found. It should be stressed, however, that these cross sections
were calculated using ``the universal'' $\sigma_{\gamma
A}(E_\gamma)/A$ dependence, which was obtained for nuclei with
$A\leq 208$.

%% Fig. 11
%%

Using the same procedure, we obtain the absolute photofission
cross section for $^{\rm nat}{\rm Pb}$ shown in
Fig.~\ref{fig:abs_bw} on a separate scale, since this cross
section was found to be much smaller than those for the actinides.
Although the results of the calculation based on set ${\cal A}$
for the $\tilde{a}_f/\tilde{a}_n$ ratios are adjusted to match the
data near the peak of $\Delta_{33}(1232)$, at higher energies they
grossly overestimate the photofission cross section. Moreover,
neglecting pre-equilibrium emission leads to a large increase of
the fission probability due to the increase of the excitation
energy of the compound nucleus. This leads to an unreasonably
large enhancement of the fission channel.

Contrary to the case for the actinides, large uncertainties in the
calculated photofission cross section for $^{\rm nat}{\rm Pb}$
make unreliable any estimate of its total photoabsorption cross
section from the photofission data~\cite{SanabriaPhd,Cetina2002}.
But since our study is aimed mainly at photofission reactions in
the actinide region, we did not attempt to obtain better agreement
with the photofission cross section for $^{\rm nat}{\rm Pb}$ via
further adjustment of the calculational parameters for each stable
isotope of lead or by introducing an empirical dependence of
fission barriers on $E^\star$, as was done in
Refs.~\cite{BAR,Sauer,Sierk}.

\subsection{Evaluation of the ``Universal Curve'' from the Actinide
Photofission Data}

It has been suggested~\cite{Frommhold1992,Frommhold1994,CLAS} that
one might be able to infer the photoabsorption cross section from
the measured photofission cross section for very heavy nuclei. In
this manner, the behavior of the ``Universal Curve'' for such
nuclei could be investigated.

\begin{figure}[ht]
\begin{centering}
{\includegraphics[width=1.0\columnwidth]{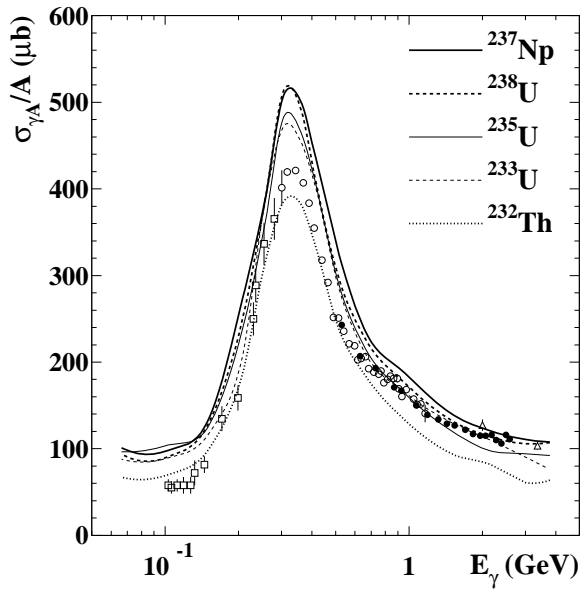}} \caption{Total
photoabsorption cross section per bound nucleon for actinide
nuclei evaluated from the photofission data of
Refs.~\cite{Sanabria,Cetina2002} and our calculated fission
probabilities. Average data for C, Al, Cu, Sn, and Pb are shown by
the open~\cite{Bianchi} and closed ~\cite{Valeria} circles.
Typical values of the systematic uncertainties for the
measurements of Ref.~\cite{Bianchi} are shown for the first and
last points. Additional experimental data for $^{238}{\rm U}$
above 2 GeV~\cite{Michalowski} and for Pb below 0.3
GeV~\cite{Lepretre,Carlos82,Vey82,Carlos84} are shown by the open
triangles and squares, respectively. } \label{fig:univ}
\end{centering}
\end{figure}

We can calculate the total photoabsorption cross section per bound
nucleon $\sigma_{\gamma A}(E_\gamma)/A$ by means of our calculated
values for $W_f(E_\gamma)$ and the actinide photofission
cross-section data of Refs.~\cite{Sanabria,Cetina2002}, using
\begin{equation}
 \frac{\sigma_{\gamma A}(E_\gamma)}{A}= \frac{\sigma_{\gamma f}(E_\gamma)}{A}\frac{1}{W_f(E_\gamma)}.
\end{equation}
The data for each of the target nuclei used in
Refs.~\cite{Sanabria,Cetina2002}, except for $^{\rm nat}{\rm Pb}$,
were interpolated by smooth curves and then divided by the
calculated $W_f$. The results of this procedure, shown in
Fig.~\ref{fig:univ}, are in agreement within $\sim 10$\% of each
other, except for $^{232}{\rm Th}$. Those for $^{232}{\rm Th}$ are
obtained with $W_f(E_\gamma)$ values that we found to be more
sensitive to the choice of calculational parameters and hence less
accurately defined. Reasonable agreement is found with the
$\sigma_{\gamma A}(E_\gamma)/A$ data of
Refs.~\cite{Bianchi,Valeria,Michalowski} at $E_\gamma\geq
0.7$~GeV, but marked disagreement in strength ($\sim 20$\%) exists
in the $\Delta_{33}$ region.

%% Fig. 12
%%

\begin{figure}[ht]
\begin{centering}
{\includegraphics[width=1.0\columnwidth]{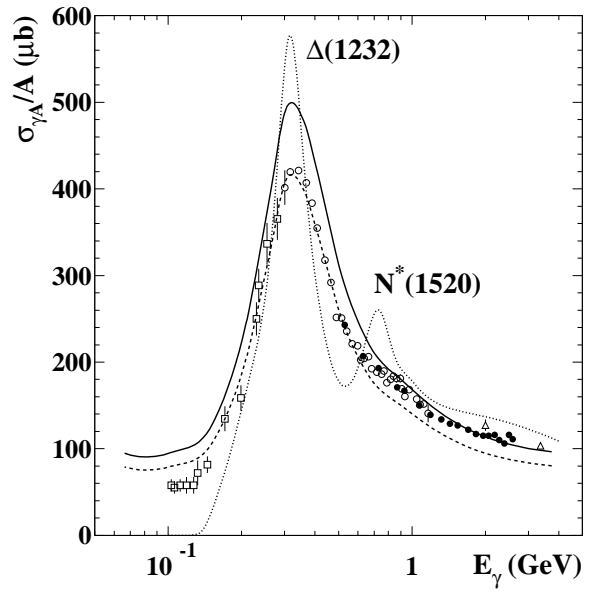}}
\caption{Average total photoabsorption cross section per bound
nucleon for $^{237}{\rm Np}$, $^{238}{\rm U}$, $^{235}{\rm U}$,
and $^{233}{\rm U}$ (solid line) using the photofission data of
Refs.~\cite{Sanabria,Cetina2002} and our calculated fission
probabilities. The dashed line represents the same estimate, but
divided by a factor of 1.2. The data points are denoted as in
Fig.~\protect\ref{fig:univ}. The total photoabsorption cross
section on a free nucleon which is obtained as a weighted average
of $\sigma_{\gamma p}$~\cite{Daresbury,DAPHNE} and $\sigma_{\gamma
n}$~\cite{Daresbury:gn} data is shown as the dotted line.}
\label{fig:univ_mean}
\end{centering}
\end{figure}

In order to understand how the mass and width of the
$\Delta_{33}(1232)$ resonance are changed in very heavy nuclei
compared with other nuclei and the nucleon, we plot
$\sigma_{\gamma A}(E_\gamma)/A$ for the average of four nuclei,
$^{237}{\rm Np}$, $^{238}{\rm U}$, $^{235}{\rm U}$, and
$^{233}{\rm U}$, as the solid curve in Fig.~\ref{fig:univ_mean}
together with the data points for nuclei with $A\leq 208$ of
Refs.~\cite{Bianchi,Valeria}. The  weighted average total
photoabsorption cross section for the free nucleon is shown as the
dotted curve:
\begin{equation}
\sigma_{\gamma N}=\frac{Z\sigma_{\gamma p}+N\sigma_{\gamma n}}{A},
\end{equation}
\noindent where $Z=92$, $N=144$, and $A=236$ were taken as average
values for these four nuclei. The total photoabsorption cross
sections on the proton $\sigma_{\gamma p}$ and on the neutron
$\sigma_{\gamma n}$ were taken as polynomial fits of experimental
data~\cite{Daresbury,DAPHNE,Daresbury:gn}. Comparison of the
nuclear data in Fig.~\ref{fig:univ_mean} with the nucleon curve
demonstrates that the $\Delta_{33}$ is shifted upward in mass
slightly and broadened considerably.

%% Fig. 13
%%

It is stated in Ref.~\cite{Bianchi} that the
$\Delta_{33}(1232)$-resonance excitation parameters increase with
nuclear density due to hadronic resonance propagation and
interactions; for Pb, compared with Li, the $\Delta_{33}$ mass
$M_\Delta$ increases  linearly up to a few tens of MeV, and its
width $\Gamma_\Delta$ up to several tens of MeV. Following this
trend, only small modifications for $M_\Delta$ and $\Gamma_\Delta$
are expected if the mass of the target nucleus increases by 30
units, from $A=208$ to 238. When the solid curve is re-scaled to
fit the experimental points close to the maximum, as shown in
Fig.~\ref{fig:univ_mean} by the dashed curve, essentially no
further modification of the mass or the width of the resonance is
required. Our study shows a far more dramatic effect: the strength
of the $\Delta_{33}(1232)$ resonance increases by $\sim 20$\%,
without any noticeable change in its mass or width.

Finally, one can also see from Figs.~\ref{fig:abs_bw},
\ref{fig:univ}, and \ref{fig:univ_mean} that a shoulder appears
for $E_\gamma\sim$~0.8--1.0~GeV. This shoulder might well reflect
the excitation of the $N^\star (1520)$ resonance. Evidence for
this $N^\star (1520)$ excitation is seen most prominently in the
structure of the JLab photofission cross sections for
$^{233,235}{\rm U}$ and $^{237}{\rm Np}$ shown in
Fig.~\ref{fig:abs_bw}.

\section{Conclusions}\label{Concl}

We have calculated, using the RELDIS model, the fission
probabilities of $^{237}{\rm Np}$, $^{233,235,238}{\rm U}$,
$^{232}{\rm Th}$, and $^{\rm nat}{\rm Pb}$ as a function of
incident photon energy. We have used these values, together with
the measured photofission cross sections from SAL and JLab, to
infer the total photoabsorption cross sections for these nuclei.
We now answer the questions raised in Sec.~\ref{SC1}.
\begin{itemize}

\item[(1)] We have been able to describe the photofission of the actinide nuclei over a broad
energy range. The calculated fission probabilities for $^{237}{\rm
Np}$, $^{238}{\rm U}$, $^{235}{\rm U}$, and $^{233}{\rm U}$ were
found to be stable within 10\% with respect to reasonable changes
of the input calculational parameters.

\item[(2)] Our model calculations show that photonuclear reactions become more complex in the GeV energy region,
when many reaction channels are open. Contrary to the naive
expectation that for the actinides, the higher the photon energy,
the closer the fission probability $W_f$ should be to unity, our
calculated fission probabilities for $E_\gamma>1$~GeV are found
not to exceed those values for $E_\gamma\sim 50-100$~MeV; they
are, in fact, predicted to decrease somewhat. At higher photon
energies, only a small part of the energy is converted into the
internal excitation of the absorbing nuclear system. The resulting
behavior of $W_f$ is due to the creation of compound nuclei with
fissility parameter $Z^2/A$ much lower than that for the target
nucleus.

\item[(3)] We show that the residual nuclei after direct and pre-equilibrium emission of hadrons
resemble the target nucleus, in the sense that the fissility
parameter $Z^2/A$ is similar, only for relatively low incident
photon energies ($E_\gamma\leq 100$ MeV). They differ markedly for
GeV energies. The products of photospallation reactions have
relatively low probabilities to undergo fission, but high rates to
decay via the less collective channels leading to the evaporation
of nucleons.

\item[(4)] Our calculations show that for the actinides, fission is not
the only significant outcome following photoabsorption at GeV
energies. Even at the highest tagged-photon energies
($E_\gamma\sim $ 4~GeV) used in the measurements at JLab, the
competition with the evaporation process is still important.
Therefore, we conclude that the photofission cross section cannot
be substituted for the total photoabsorption cross section in the
GeV energy region, even for those nuclei with the largest
fissility parameters. The photoabsorption cross section must be
obtained from a calculation of $W_f$ applied to the photofission
cross section.

\item[(5)] The total photoabsorption cross sections per bound nucleon for
$^{237}{\rm Np}$ and $^{233,235,238}{\rm U}$ were inferred using
the recent photofission data~\cite{Sanabria,Cetina2002} and our
calculated fission probabilities. The resulting ``universal''
curves for $\sigma_{\gamma A}(E_\gamma)/A$ were found to be in
agreement within $\sim 10$\% of each other. However, reasonable
agreement with $\sigma_{\gamma A}(E_\gamma)/A$ obtained for nuclei
with $A\leq 208$ was found only for $E_\gamma\geq 0.8$~GeV, while
our values for the heavy actinides are $\sim 20$\% higher in the
region of the $\Delta_{33}(1232)$ resonance.

\item[(6)] The calculational uncertainties for fission probabilities
for Np and U are estimated to be at the level of 10\%, which
should be combined with the $\sim 5$\% accuracy of the measured
photofission cross sections. Although the disagreement found with
$\sigma_{\gamma A}(E_\gamma)/A$ for $A\leq 208$ nuclei cannot
readily be explained by the presence of these uncertainties, one
also should take into account the accuracy of previous
measurements of the ``Universal Curve,'' which can be estimated as
no better than 5\% as well. These uncertainties taken together
prevent us from drawing a definitive conclusion that the
``Universal Curve,'' obtained for nuclei with $A\leq 208$, breaks
down for heavy nuclei with $A\geq 233$ and for $E_\gamma$ in the
$\Delta_{33}(1232)$-resonance region. However, our findings
clearly call into question this concept and demonstrate the need
for direct measurement of the total photoabsorption cross sections
for the heavy actinides. Such measurements would serve to confirm
or refute the predictions of this paper.

\end{itemize}

\begin{acknowledgments}

I.A.P. thanks the Center for Nuclear Studies of The George
Washington University for warm hospitality and financial support
and the Russian Foundation for Basic Research for support in
part under Grant RFFI-02-02-16013. This work is supported by the
US Department of Energy under Grant  DE-FG02-95ER40901.
\end{acknowledgments}

%%%%%%%%%%%%%%%%%%%%%%%%%%%%%%%%
\end{document}